\documentclass[acmsmall]{acmart}

\usepackage{tikz}
\usetikzlibrary{arrows, shapes, positioning, shadows, trees}
\tikzset{
  basic/.style  = {draw, text width=2cm, drop shadow, font=\sffamily, rectangle},
  root/.style   = {basic, rounded corners=2pt, thin, align=center, text width = 11em, fill=green!30},
  level 2/.style = {basic, rounded corners=6pt, thin,align=center, fill=green!60,
                  text width=11em},
  level 3/.style = {basic, thin, align=left, fill=pink!60, text width=12.5em}
}

\usepackage[super]{nth}
\usepackage{amsmath}
\usepackage{amsthm}
\usepackage{bm}
\usepackage{array}
\usepackage{makecell}
\usepackage{multirow}
\usepackage{tabularx}
\usepackage{paralist}
\usepackage{graphicx}
\usepackage{caption}
\usepackage{subcaption}
\usepackage{url}
\usepackage{xcolor}
\usepackage{enumitem}
\usepackage{float}
\usepackage[flushleft]{threeparttable}
\usepackage[ruled,vlined,linesnumbered]{algorithm2e}
\usepackage{multirow}
\usepackage{multicol}

\usepackage{pifont}
\newcommand{\cmark}{\ding{51}}%
\newcommand{\xmark}{\ding{55}}%

\makeatletter
\newcommand{\mathleft}{\@fleqntrue\@mathmargin0pt}

\theoremstyle{definition}
\newtheorem{definition}{Definition}[section]
\usepackage{bbm}

\newcounter{todocounter}
\AtBeginDocument{%
  \providecommand\BibTeX{{%
    \normalfont B\kern-0.5em{\scshape i\kern-0.25em b}\kern-0.8em\TeX}}}

\begin{document}
\setlength{\abovedisplayskip}{3.75pt plus 0pt}%
\setlength{\belowdisplayskip}{3.75pt plus 0pt}%
\title{Fairness and Diversity in Recommender Systems: A Survey}
\renewcommand{\shorttitle}{Fairness and Diversity in Recommender Systems}

\author{Yuying Zhao}
\email{yuying.zhao@vanderbilt.edu}
\author{Yu Wang}
\email{yu.wang.1@vanderbilt.edu}
\author{Yunchao Liu}
\email{yunchao.liu@vanderbilt.edu}
\author{Xueqi Cheng}
\email{xueqi.cheng@vanderbilt.edu}
\affiliation{%
  \institution{Vanderbilt University}
  \country{USA}
}
\author{Charu C. Aggarwal}
\email{charu@us.ibm.com}
\affiliation{%
  \institution{IBM T.J. Watson Research Center}
  \country{USA}
}
\author{Tyler Derr}
\email{tyler.derr@vanderbilt.edu}
\affiliation{%
  \institution{Vanderbilt University}
  \country{USA}
}




\begin{abstract}
Recommender systems (RS) are effective tools for mitigating information overload and have seen extensive applications across various domains. However, the single focus on utility goals proves to be inadequate in addressing real-world concerns, leading to increasing attention to fairness-aware and diversity-aware RS. While most existing studies explore fairness and diversity independently, we identify strong connections between these two domains. In this survey, we first discuss each of them individually and then dive into their connections. Additionally, motivated by the concepts of user-level and item-level fairness, we broaden the understanding of diversity to encompass not only the item level but also the user level. With this expanded perspective on user and item-level diversity, we re-interpret fairness studies from the viewpoint of diversity. This fresh perspective enhances our understanding of fairness-related work and paves the way for potential future research directions. Papers discussed in this survey along with public code links are available at: \href{https://github.com/YuyingZhao/Awesome-Fairness-and-Diversity-Papers-in-Recommender-Systems}{\textcolor{blue}{https://github.com/YuyingZhao/Awesome-Fairness-and-Diversity-Papers-in-Recommender-Systems}}.

\end{abstract}


\maketitle

\section{Introduction}

To tackle the challenges of information overload~\cite{aljukhadar2012using}, recommender systems are playing a crucial role in providing personalized services to fit users' interests. Their effectiveness has been demonstrated across various applications, including news recommendations~\cite{liu2010personalized}, product recommendations~\cite{pal2020pinnersage, he2023dynamically}, friend recommendations~\cite{xie2010potential, fan2019graph}, and crystal recommendations~\cite{pandey2021predicting, qu2023leveraging}. These systems not only improve users' experience but also increase entity exposure, which thereby boosts the profits of content providers. The primary goal of these systems is to improve utility performance (e.g., recall, click-through rate)~\cite{aggarwal2016recommender}. However, solely pursuing this goal may lead to practical issues (e.g., Mattew Effect~\cite{merton1968matthew}, Filter Bubble~\cite{nguyen2014exploring}, etc). Consequently, researchers have considered other aspects, such as fairness~\cite{li2022fairness}, diversity~\cite{kunaver2017diversity}, explainability~\cite{chen2022grease}, privacy~\cite{jeckmans2013privacy}, robustness~\cite{chen2023dark, li2022towards}, long-term benefits~\cite{chen2023sim2rec}, etc. Acknowledging the significance of beyond-utility perspectives, this survey provides an in-depth discussion of fairness and diversity in RS. 

Fairness and diversity are of great importance in RS~\cite{li2022fairness, wang2023survey, kaminskas2016diversity, kunaver2017diversity}. Studies have revealed that RS can potentially exhibit unfairness, adversely affecting the interests of multiple stakeholders~\cite{abdollahpouri2019multi,ge2021towards}. Given the increasing societal influence, any biases embedded within these systems have significant impacts. For example, if popular items from big companies dominate the recommendations, the development of small businesses will be hindered, magnifying economic disparities and adversely affecting the health of business ecosystem. To address these growing concerns, fairness-aware RS have gained increasing attention, which have been thoroughly investigated from both user level and item level. User-level fairness~\cite{li2021user,burke2018balanced,wu2021learning} seeks to ensure equitable treatment across different user groups (e.g., groups based on gender, race, etc), while item-level fairness~\cite{beutel2019fairness,nandy2022achieving,ge2021towards} requires that different item groups (e.g., popular and unpopular items) will have equal opportunities of being recommended. In addition to fairness, considering diversity is also conducive to RS. Researchers have shown that without diversity consideration (i.e., only focusing on relevance scores), RS tend to recommend homogeneous/similar items~\cite{wu2022survey}, which may potentially harm both customers (user side) and providers (item side). For customers, a proliferation of similar/redundant items can lead to user fatigue/boredom and decreased long-term satisfaction~\cite{mcnee2006being}. For providers, small businesses might suffer from low exposure~\cite{rahmani2022unfairness} due to the dominance of large companies in the market.

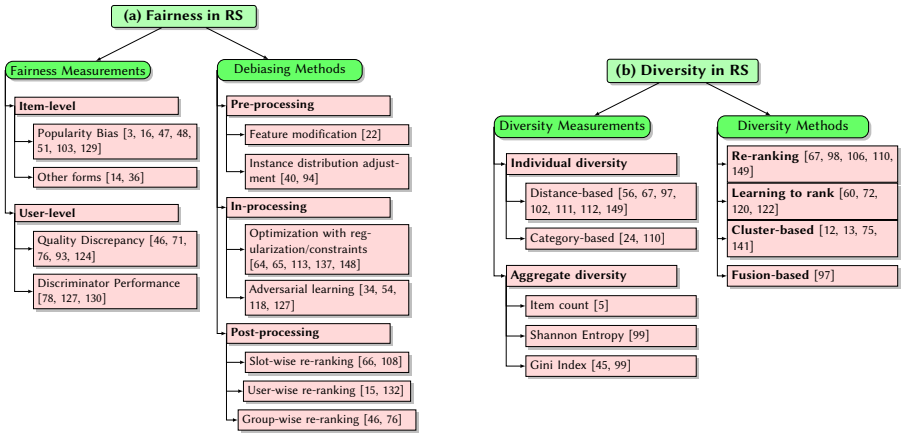
\begin{figure}

     \centering
     \hspace{-1ex}
     \begin{subfigure}[c]{0.3\textwidth}
          \hspace{-10ex}
         \centering
\scalebox{0.47}
{
\begin{tikzpicture}[
  level 1/.style={sibling distance=60mm},
  edge from parent/.style={->,draw},
  >=latex]

 \node[root] {\Large{\textbf{(a) Fairness in RS}}}
  child {node[level 2] (c1) {\large Fairness Measurements}}
  child {node[level 2] (c2) {\large Debiasing Methods}};

\begin{scope}[every node/.style={level 3}]
\node [below of = c1, xshift=15pt] (c11) {\textbf{Item-level}};
\node [below of = c11, xshift=15pt] (p11) {Popularity Bias \cite{abdollahpouri2019unfairness,singh2018fairness, ge2022explainable, ge2021towards,borges2021mitigating, wu2021tfrom,gomez2021winner}};
\node [below of = p11] (p12) {Other forms \cite{beutel2019fairness,deldjoo2021flexible}};

\node [below of = p12, xshift=-15pt] (c12) {\textbf{User-level}};
\node [below of = c12, xshift=15pt] (p21) {Quality Discrepancy \cite{fu2020fairness,li2021user,rahmani2022experiments,wu2022big,leonhardt2018user}};
\node [below of = p21, yshift = -5.5pt,] (p22) {Discriminator Performance \cite{wu2021learning,wu2022selective,li2021towards}};

\node [below of = c2, xshift=15pt] (c21) {\textbf{Pre-processing}};
\node [below of = c21, yshift = 5.5pt, xshift=15pt] (d11) {Feature modification \cite{chencounterfactual}};
\node [below of = d11, yshift = -1pt] (d12) {Instance distribution adjustment \cite{rastegarpanah2019fighting,ekstrand2018all}};

\node [below of = d12, xshift = -15pt] (c22) {\textbf{In-processing}};
\node [below of = c22, xshift=15pt, yshift=-5pt] (d21) {Optimization with regularization/constraints \cite{kamishima2013efficiency,kamishima2018recommendation,yao2017beyond,wan2020addressing,zhu2018fairness}};
\node [below of = d21, yshift=-10pt] (d22) {Adversarial learning \cite{wu2021learning, wang2022improving,dai2021say, grari2023adversarial}};

\node [below of = d22, xshift = -15pt] (c23) {\textbf{Post-processing}};
\node [below of = c23, yshift = 5.5pt, xshift=15pt] (d31) {Slot-wise re-ranking \cite{steck2018calibrated,karako2018using}};
\node [below of = d31, yshift = 5.5pt, text width=4.5cm] (d32) {User-wise re-ranking \cite{xiao2017fairness,biega2018equity}};
\node [below of = d32, yshift = 5.5pt, text width=4.8cm] (d33) {Group-wise re-ranking \cite{fu2020fairness,li2021user}};

\end{scope}

\foreach \value in {1,2}
  \draw[->] (c1.west) |- (c1\value.west);

\foreach \value in {1,2,3}
  \draw[->] (c2.west) |- (c2\value.west);


  \draw[->] (c11.west) |- (p11.west);
  \draw[->] (c11.west) |- (p12.west);

  \draw[->] (c12.west) |- (p21.west);
  \draw[->] (c12.west) |- (p22.west);

  
\foreach \value in {1, 2}
  \draw[->] (c21.west) |- (d1\value.west);

  \draw[->] (c22.west) |- (d21.west);

  \draw[->] (c22.west) |- (d22.west);

  \draw[->] (c23.west) |- (d31.west);
  \draw[->] (c23.west) |- (d32.west);
  \draw[->] (c23.west) |- (d33.west);

\end{tikzpicture}
}
\label{fig: fairness_tree}
     \end{subfigure}
     \hspace{5ex}
     \begin{subfigure}[c]{0.3\textwidth}
         \centering
\scalebox{0.49}
{
\begin{tikzpicture}[
  level 1/.style={sibling distance=60mm},
  edge from parent/.style={->,draw},
  >=latex]

\node[root] {\Large{\textbf{(b) Diversity in RS}}}
  child {node[level 2,text width=4cm] (c1) {\large Diversity Measurements}}
  child {node[level 2] (c2) {\large Diversity Methods}};

\begin{scope}[every node/.style={level 3}]
\node [below of = c1, xshift=15pt] (c11) {\textbf{Individual diversity}};
\node [below of = c11, xshift=15pt] (p11) {Distance-based \cite{ziegler2005improving,kelly2006enhancing, vargas2011intent, shi2012adaptive,ribeiro2012pareto,vargas2011rank,haritsa2009kndn}};
\node [below of = p11] (p12) {Category-based \cite{vargas2014coverage,chen2007preference}};

\node [below of = p12, xshift=-15pt] (c12) {\textbf{Aggregate diversity}};
\node [below of = c12, yshift = 5.5pt, xshift=15pt] (p21) {Item count \cite{adomavicius2011improving}};
\node [below of = p21, yshift = 5.5pt ] (p22) {Shannon Entropy \cite{shani2011evaluating}};
\node [below of = p22, yshift = 5.5pt ] (p23) {Gini Index \cite{fleder2007recommender,shani2011evaluating}};

\node [below of = c2, xshift=15pt] (c21) {\textbf{Re-ranking} \cite{smyth2001similarity, ziegler2005improving, kelly2006enhancing, vargas2014coverage, santos2010exploiting}};
\node [below of = c21] (c22) {\textbf{Learning to rank} \cite{wasilewski2016incorporating,hurley2013personalised,wang2023diversity,li2020cascading}};
\node [below of = c22] (c23) {\textbf{Cluster-based} \cite{zhang2009novel, aytekin2014clustering, berbague2021overlapping, li2012multidimensional}};
\node [below of = c23] (c24) {\textbf{Fusion-based} \cite{ribeiro2012pareto}};

\end{scope}

\foreach \value in {1,2}
  \draw[->] (c1.west) |- (c1\value.west);

\foreach \value in {1,2,3,4}
  \draw[->] (c2.west) |- (c2\value.west);

  \draw[->] (c11.west) |- (p11.west);
  \draw[->] (c11.west) |- (p12.west);

  \draw[->] (c12.west) |- (p21.west);
  \draw[->] (c12.west) |- (p22.west);
  \draw[->] (c12.west) |- (p23.west);

  
\end{tikzpicture}
}
\label{fig: diversity_tree}
     \end{subfigure}
     \vskip -3.5ex
        \caption{(a) Fairness in Recommender Systems: fairness measurements and debiasing methods. (b) Diversity in Recommender Systems: diversity measurements and methods to enhance diversity. 
        }

\label{fig:measurements-and-methods}
\end{figure}

\begin{figure}
    \centering
    \includegraphics[width=0.98\textwidth]{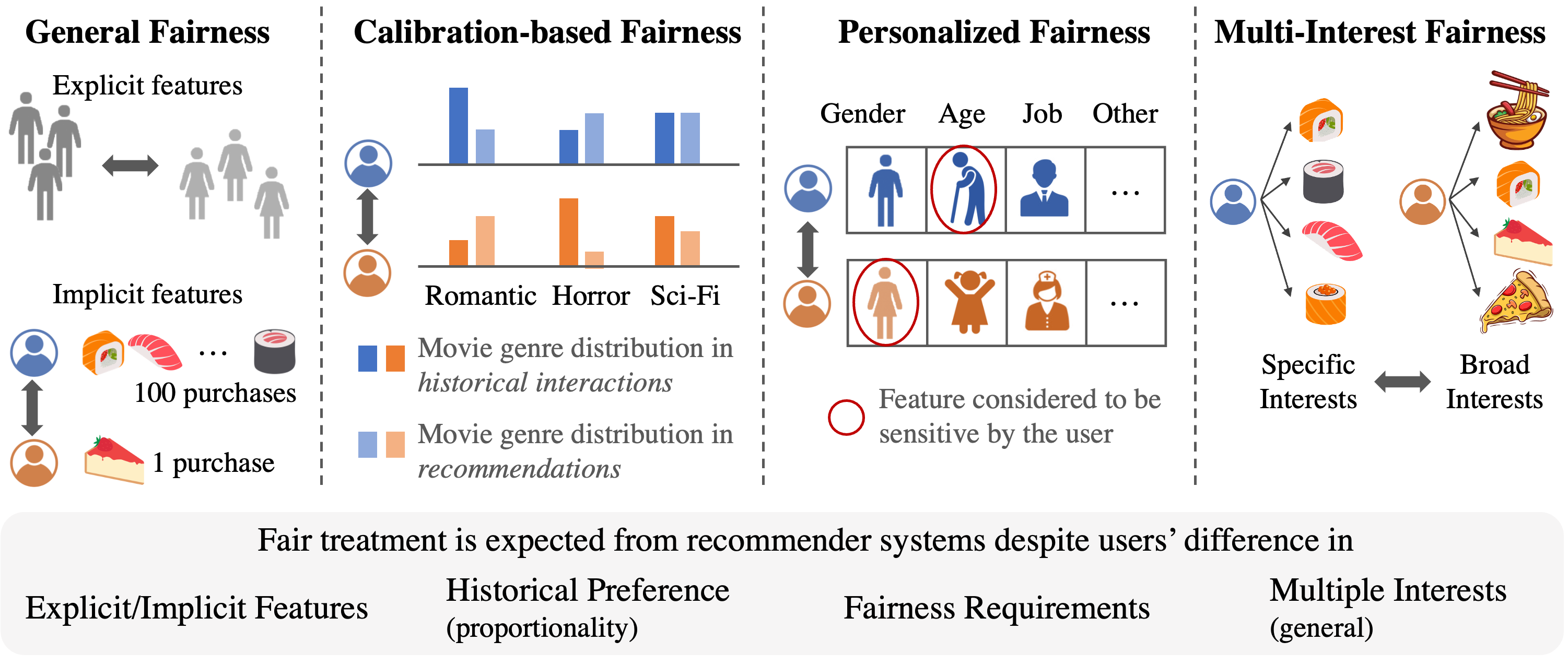}
    \caption{Fairness and Diversity in RS: users are expected to be treated fairly despite their differences.}
    \label{fig:fairness_and_diversity_visualization}
\end{figure}

Although fairness and diversity have been exhaustively investigated independently with measurements and methods summarized in Fig.~\ref{fig:measurements-and-methods}(a)(b), their intrinsic connection remains insufficiently explored. The examination of fairness-diversity relationship presents several benefits, including an enhanced comprehension of the intersection and the revelation of potential research directions which are often overlooked when these domains are studied in isolation. To address this oversight, our analysis encompasses both item and user levels, with an emphasis on the latter, an aspect often underrepresented in diversity research. While both fairness and diversity bear significant implications for users and items alike, discussions on fairness are commonly conducted from both perspectives, whereas diversity tends to be primarily examined in the context of items. Thus, it becomes imperative to expand the scope of diversity to incorporate user aspects as well. We categorize user diversity into \textit{explicit/implicit features, historical preferences (proportionality), fairness requirements, and multiple interests (general)} as shown in Fig.~\ref{fig:fairness_and_diversity_visualization}. With these expanded diversity definitions, fairness works can be re-interpreted from diversity perspective at both user and item levels. In terms of fairness-diversity connection at the user level, strategies that promote fairness can be construed as mechanisms to alleviate disparate treatment of users, grouped based on different diversity metrics. We further elucidate this by providing a comprehensive table encapsulating these works from the perspective of user diversity. From an item-level standpoint, augmenting item diversity serves as an efficacious strategy for promoting item fairness~\cite{sonboli2020opportunistic, liu2019personalized}. We also conduct the experiment to empirically explore the relationship between fairness and diversity.

In conclusion, this survey delivers several key contributions: First, we propose a novel categorization of user diversity, thereby expanding the conventional conceptualization of diversity focusing on the item side. Second, we delve into an exhaustive discussion of fairness-diversity connection at both user and item levels. Our exploration reveals that fairness works can be re-interpreted through the lens of diversity, and strategies enhancing diversity have proven efficacious in improving fairness. Additionally, we delineate pertinent concepts within fairness and diversity individually, advancing existing surveys with more recent works and ensuring the audience is adequately equipped with the requisite contextual understanding prior to delving into their connections.

\textbf{Relations to other surveys:} 
Various surveys have been published in recent years focusing either on fairness or diversity. Although some of them mention briefly the other aspect (i.e., discuss diversity in fairness surveys~\cite{deldjoo2022survey} or discuss fairness in diversity surveys~\cite{wu2022survey}), none of them have comprehensively discussed the connection between these two domains. In this survey, we aim to fill this crucial gap by focusing on the connections in addition to covering them individually to provide the context. Our aim is not to offer exhaustive discussions on single aspects, as these have been covered in existing surveys for fairness~\cite{wang2023survey, deldjoo2022survey,li2022fairness,chen2023bias} and diversity~\cite{wu2022survey,wu2019recent,kaminskas2016diversity,kunaver2017diversity,castells2021novelty}. Rather, our goal is twofold: firstly, for both fairness and diversity individually, we will augment existing knowledge with more recent developments, recognizing the rapidly expanding body of work in these areas; secondly, we will provide a thorough discussion and categorization of the connection between fairness and diversity. For fairness, we focus on the user and item side. While other more complex categorizations exist (e.g., single-side versus multi-side, dynamic versus static), they are not the major focus of this paper. Regarding diversity, previous works generally discussed it from the item side, while in this survey, we also discuss diversity from the novel user side. The discussions provide new perspectives on understanding fairness from the view of diversity. 

\textbf{Paper organization:} The rest of this survey is organized as follows. In Section~\ref{sec.preliminary}, we introduce RS preliminaries. Next, we discuss fairness and diversity individually in Section~\ref{sec-fairness} and Section~\ref{sec-diversity}. Then, in Section~\ref{sec-fd}, we discuss the fairness-diversity connection and extend diversity concepts to user side motivated by the categorization of the fairness domain. Based on the diversity concepts including user-side and item-side, we review the fairness works based on each diversity aspect. We also conduct experiments to empircally investigate the trade-off between fairness and diversity. In the end, we discuss challenges and opportunities in Section~\ref{sec-challenges} and conclude the survey in Section~\ref{sec-conclusion}.

\section{Recommender System Preliminaries}
\label{sec.preliminary}

RS are designed to mitigate information overload by recommending items that match users' interests. A typical RS consists of a user set $\mathcal{U}=\{u_1, u_2, ..., u_n\}$ with $n$ users and an item set $\mathcal{I}=\{i_1, i_2, ..., i_m\}$ with $m$ items. The user-item historical interactions are represented by a matrix $\mathbf{A}\in\mathbb{R}^{n\times m}$ where $\mathbf{A}_{ui}$ denotes whether user $u$ has interacted with item $i$. We note that in most cases $\mathbf{A}_{ui}\in\{0, 1\}$ for binary interactions but could also be weighted, e.g., rating score and purchasing number. The primary goal of recommendations is to predict the list of top $K$ items for each user given the historical user-item interactions $\mathbf{A}$. The top $K$ items are selected based on the preference/relevance scores, which are the dot products between user embedding $\mathbf{E}_{\mathcal{U}}$ and item embedding $\mathbf{E}_{\mathcal{I}}$. For instance, the relevance score between user-item pair $(u, i)$ is ${\mathbf{e}_u}^\top \mathbf{e}_i$ where $\mathbf{e}_u$ and $\mathbf{e}_i$ are the embeddings of user $u$ and item $i$. After obtaining the scores, $K$ items with the highest values are recommended. To learn the representations for preference calculation, RS can be divided into collaborative filtering (CF)-based and content-based. CF-based methods~\cite{wang2022collaboration, aggarwal2016neighborhood} utilize the user-item interactions and recommend items based on users with similar interaction patterns. Content-based methods~\cite{adomavicius2010context,aggarwal2016content}, on the other hand, usually use additional features (e.g., user profiles) to assist the recommendation process. We direct interested readers to existing RS surveys/books~\cite{aggarwal2016recommender,zhang2019deep} for further details.

Formally, the problem definition of standard recommendation is established as follows: given the interaction matrix $\mathbf{A}\in \mathbb{R}^{n\times m}$, the goal is to learn the function $f: \mathcal{U}\times \mathcal{I}\rightarrow \mathbf{R}$, such that the predicted preference matrix $\mathbf{R}$ approximates the true (including unobserved) preference of the users for the items as closely as possible. For top $K$ recommendations, the system recommends the top $K$ items to each user, with these items having highest scores, and formalized as $\mathcal{R}_u^K=\{i|i\in \mathcal{I} \And \mathbf{R}_{ui}\in \text{top} K(\mathbf{R}_u)\}$. The notations in this survey are summarized in Table~\ref{tab:notations}. Regarding user and item embeddings, for simplicity, we use $\mathbf{e}_u$ and $\mathbf{e}_i$ to denote user and item embeddings if no confusion. Otherwise, we will use $\mathbf{e}^U_i$ and $\mathbf{e}^I_i$ to distinguish user and item embeddings with the superscripts.

\begin{table}[t]
\footnotesize 
    \centering
    \caption{Common notations used throughout the survey and their associated descriptions.}
    \vspace{-2ex}
    \begin{tabular}{|cc|cc|}
    \hline
        Notation & Description & Notation & Description \\
        \hline
        $K$ & The number of recommended items & $\mathbf{A}/\mathbf{R}$ & Interaction/Predicted preference matrix \\
         $\lambda$ & Coefficient & $\mathbf{W}$ & Binary matrix of recommended items\\
         $i^*$ & Selected item to be added into recommendation & $\mathbf{E}_U$/$\mathbf{e}_u$ & User embeddings/user $u$'s embedding \\
         $K^\prime$ & The number of user interests & $\mathbf{E}_I$/$\mathbf{e}_i$ & Item embeddings/item $i$'s embedding  \\
         $a$ & Attention & $\mathbf{Z}_u$ & $u$'s representation with multiple interests\\
         \hline
         $\mathcal{U}$ & User set with $n$ users  & $Q(\cdot)$ & Recommendation quality function \\  
         $\mathcal{I}$ & Item set with $m$ items  & $d(\cdot, \cdot)$ & Distance function \\
         $\mathcal{R}_u^K$/$\mathcal{R}_u$ & Top $K$ recommendation list for user $u$ & $f_{\text{rec}}(\cdot)$ & Function to measure utility performance\\
         $\mathcal{G}_i$ & $i$-th user/item group  & $f_{\text{fair}}(\cdot)$ & Function to measure fairness performance\\
         $\mathcal{C}$ & Item candidate set (re-ranking)  & $f_{\text{div}}(\cdot)$ & Function to measure diversity performance\\
         $\mathcal{S}$ & (In)complete 
         recommendation set (re-ranking)  & $\mathcal{L}_{\text{rec}}$ & Loss term for utility performance \\
         $\mathcal{D}$ & Provider set & $\mathcal{L}_{\text{fairness}}$ & Loss term for fairness performance \\
         \hline
    \end{tabular}
    \label{tab:notations}
    \vskip -2ex
\end{table}

\vspace{-1ex}
\section{Fairness in Recommender Systems}
\label{sec-fairness}
In this section, we introduce the fairness measurements to quantify fairness and debiasing methods to enhance fairness. A summary is shown in Figure~\ref{fig:measurements-and-methods} (a).

\vspace{-1ex}
\subsection{Fairness Measurements}
As one of the most representative multi-stakeholder systems, RS raise fairness concerns from both item (i.e., item-level fairness) and user (i.e., user-level fairness) sides. While other categories exist (e.g., group vs individual), we refer readers to other survey~\cite{li2022fairness,wang2023survey} for a comprehensive discussion. In this survey, we focus on user-level and item-level fairness, given their innate links to diversity.

\subsubsection{Item-Level Fairness}
Item-level fairness focuses on the fair treatment of items during recommendations. One of the most predominant concerns in RS, particularly at the item level, is popularity bias~\cite{abdollahpouri2019unfairness}, where generally RS tend to recommend popular items to users. Popularity bias would lead to exposure unfairness. In this context, exposure refers to the chance of an item being recommended which is measured by its occurrence in the top $K$ recommendation. Consequently, popular items receive more exposure and thereby gain more popularity, widening the disparity between popular and less-popular items. This will be detrimental to users, the provider selling the items, and the platform~\cite{abdollahpouri2017controlling,ekstrand2018all}:
(1) Users' preference towards unpopular items would be under-represented due to the majority training towards popular items; (2) It becomes hard for the growth of small businesses even if they can provide items with similar quality with popular items; (3) If the producers cannot sell products and make benefits, highly likely they will leave the platform, which may inform corporate monopoly and is unhealthy for the platform development in the long run. Typically, exposure fairness measures the difference in exposure for the items belonging to distinct groups, which is defined based on various constraints, including demographic parity constraints~\cite{singh2018fairness, ge2022explainable, ge2021towards} and Extract-$K$ fairness constraint~\cite{ge2021towards}. Next, we first give the formal definition of exposure and define fairness metrics based on exposure function. Note that other functions can also be used to calculate exposure. Interested readers could refer to paper~\cite{deldjoo2023fairness} where they provide a summary table and detailed exposure metrics and corresponding fairness metrics. 

\begin{definition}
    \textbf{Exposure} measures item occurrences in users' recommendation. If an item is recommended to more users, this item has a higher exposure.
    \begin{equation}
    \small 
        \text{Exposure}(i) = \sum_{u \in \mathcal{U}}\mathbbm{1}(i \in \mathcal{R}_u).
    \end{equation}
\end{definition}

\begin{definition}
    \textbf{Demographic parity-based exposure fairness} requires the average exposure of item groups to be equal. It is defined as: 
    \begin{equation}
    \small 
    \Bigg|\frac{1}{|\mathcal{G}_1|}\sum_{i\in \mathcal{G}_1}\text{Exposure}(i)=\frac{1}{|\mathcal{G}_2|}\sum_{i\in \mathcal{G}_2}\text{Exposure}(i)\Bigg|,
    \label{eq.demographic}
\end{equation}
where $\mathcal{G}_1$ and $\mathcal{G}_2$ are groups divided by item's popularity.
\end{definition}

To allow a flexible adjustment in practice,  Extract-$K$ fairness constraint~\cite{ge2021towards} introduces $\alpha$.
\begin{definition}
    \textbf{Extract-$K$-based exposure fairness} requires the exposure of various groups are statistically indistinguishable from a given maximum $\alpha$:
    \begin{equation}
    \small 
    \frac{\sum_{u\in \mathcal{G}_1}\text{Exposure}(i)}{\sum_{i\in \mathcal{G}_2}\text{Exposure}(i)}=\alpha
    \label{eq.extract_k}
\end{equation}
\end{definition}
When $\alpha=\frac{|\mathcal{G}_1|}{|\mathcal{G}_1|}$, Eq.~(\ref{eq.extract_k}) equals Eq.~(\ref{eq.demographic}). While Eq.~(\ref{eq.extract_k}) and Eq.~(\ref{eq.demographic}) are strictly fair, it is challenging to achieve this goal in practice. Therefore, exposure fairness is often defined in the disparity form.

\begin{definition}
    \textbf{Disparity-based exposure fairness} measures the difference between exposures:
\begin{equation*}
\scriptsize  
    F(\mathcal{G}_1, \mathcal{G}_2)= \Bigg| \frac{1}{|\mathcal{G}_1|}\sum_{i\in \mathcal{G}_1}\text{Exposure}(i)-\frac{1}{|\mathcal{G}_2|}\sum_{i\in \mathcal{G}_2}\text{Exposure}(i)\Bigg| \text{~}, \quad\text{ \small or }
    =\bigg|\sum_{i\in \mathcal{G}_1}\text{Exposure}(i)-\alpha\sum_{i\in \mathcal{G}_2}\text{Exposure}(i)\bigg|
\end{equation*}
\end{definition}

A lower score indicates a higher level of fairness. We note other forms have been studied to measure popularity bias~\cite{borges2021mitigating, wu2021tfrom,gomez2021winner} and item fairness besides popularity bias~\cite{beutel2019fairness,deldjoo2021flexible}. Additionally, the above exposure fairness definitions are for two groups. If there are more than two groups, the definitions can be extended to consider the disparities across all pairs of groups.

\subsubsection{User-Level Fairness}
There are two directions of user-level fairness. One is based on recommendation quality discrepancy, and the other is whether the recommendation encodes sensitive features. We denote the first as \textit{Quality Discrepancy} and the second as \textit{Discriminator Performance} as whether encoding sensitive features is commonly measured with discriminator performance.

\noindent\textbf{Quality Discrepancy:} Fairness on the user side is related to the recommendation quality, which can be defined as the performance gap among groups and Gini coefficient~\cite{gini1921measurement}.

\begin{definition}
\textbf{Performance unfairness (group level)} measures the gap in recommendation performance between groups.
    \begin{equation}
    \small 
    \text{Unfairness}_\text{Gap}(\mathcal{G}_1, \mathcal{G}_2, \mathbf{R})=\Bigg|
    \frac{1}{|\mathcal{G}_1|}\sum_{u\in \mathcal{G}_1}Q(\mathcal{R}_u)
    -
    \frac{1}{|\mathcal{G}_2|}\sum_{u\in \mathcal{G}_2}Q(\mathcal{R}_u)
    \Bigg| \text{~},
    \label{eq.gap_unfairness}
\end{equation}
where $\mathbf{R}$ is the recommendations obtained from a recommender system with $\mathcal{R}_u$ denoting the top $K$ recommendation list for user $u$, and $Q$ is a general quality measurement (e.g., F1, NDCG).
\end{definition}
Under this framework, various works define different quality measurements and leveraging different group partition strategies. Fu et al. ~\cite{fu2020fairness} investigate the unfairness between active users and inactive users based on the number of purchases. They use F1 and NDCG as recommendation quality metrics and also propose explanation quality metrics related to the diversity of explainable paths in the knowledge graph as another quality measurement. Li et al.~\cite{li2021user} divide users into the advantaged and disadvantaged groups according to multiple criteria including interaction number, total consumption, and max prize. They use F1 and NDCG as the quality metrics. Rahmani et al.~\cite{rahmani2022experiments} conduct comprehensive experiments on various domains and datasets based on \cite{li2021user}. In addition to the groups divided by the level of activity (i.e., interaction number), they also investigate the unfairness between advantaged and disadvantaged groups based on the consumption of popular items.
Wu et al.~\cite{wu2022big} investigate the unfairness between cold and heavy users which are divided by the number of historical news clicks. They use AUC as the quality measurement. They also define unfairness based on performance gap. However, different from Eq.~(\ref{eq.gap_unfairness}), they obtain the optimal checkpoints performance for all users and cold users respectively and measure the gap.

\begin{definition}
    \textbf{Performance unfairness (individual level)} measures pairwise performance disparity between instances without group concept. It is defined based on Gini coefficient where the pairwise disparity between two users is normalized by the average performance:
    \begin{equation}
    \small 
    \text{Unfairness}_{Gini}(\mathbf{R})=\frac{\sum_{u_1, u_2\in \mathcal{U}}|Q(\mathcal{R}_{u_1})-Q(\mathcal{R}_{u_2})|}{2|\mathcal{U}|\sum_{u\in \mathcal{U}}Q(\mathcal{R}_u)}.
    \label{eq.gini_unfairness}
\end{equation}

\end{definition}
Similar to the group performance gap mentioned above, Fu et al.~\cite{fu2020fairness} define the quality as regular NDCG and F1 and also the explanation quality as $Q$ in Eq.~(\ref{eq.gini_unfairness}). Leonhardt et al.~\cite{leonhardt2018user} define $Q$ based on the preference scores of the recommended items and their top $K$ items.

\noindent\textbf{Discriminator Performance:} RS aims to learn high-quality user representations which encode users' preferences for downstream recommendations. It is critical to investigate whether the learned embeddings are fair. This involves sensitive features (e.g., gender) and a discriminator trained to predict sensitive features given user embeddings. If the performance (e.g., accuracy) is low for the discriminator, the recommendation model satisfies the fairness requirement in the embedding space, indicating a high level of fairness. Wu et al.~\cite{wu2021learning} study gender bias and use AUC to measure the discriminator performance of binary classification to mitigate the impact of data imbalance. When the sensitive attribute is not binary (i.e., multiple values), they use micro-averaged F1 measure. Similarly, Wu et al.~\cite{wu2022selective} use F1 score as the measurement for the discriminator performance, where users are allowed to choose sensitive features (i.e., personalized fairness). Li et al.~\cite{li2021towards} also study personalized fairness and use AUC for discriminator performance.

\vspace{-1ex}
\subsection{Debiasing Methods}
\label{sec.debias}
Numerous efforts have been devoted to designing debiasing methods~\cite{li2022fairness,wang2023survey}. According to the phase of the intervention, they can be summarized into (1) \textit{pre-processing methods} which debias the data before training; (2) \textit{in-processing methods} which incorporate fairness consideration into training process; (3) \textit{post-processing methods} which adjust the recommendation after the model is trained. These methods can be used individually or simultaneously in different phases.

\subsubsection{Pre-processing Methods}
Pre-processing methods aim to adjust the training data so that it contains less bias. Therefore, the model trained on unbiased data would be fairer. To achieve this, one can either modify the features of the training data~\cite{chencounterfactual} or change the instance distribution (e.g., delete, add, re-sample)~\cite{rastegarpanah2019fighting,ekstrand2018all} without feature update. One naive way, which falls in the category of suppression, is to remove sensitive features from input features~\cite{kusner2017counterfactual}.
However, on the one hand, simply removing features might hurt recommendation performance. On the other hand, features are correlated, and thus removing the sensitive features cannot guarantee fairness~\cite{wang2022improving}. Therefore, other ways are developed based on orthogonalization~\cite{chencounterfactual} and marginal distribution mapping~\cite{chencounterfactual}. Additionally, adversarial approaches are used to learn fair features but they are more commonly twined with the downstream tasks. Therefore, we categorize them into in-processing methods. Additionally, instances can be adjusted in the training set without changing the features. For example, Rastegarpanah et al.~\cite{rastegarpanah2019fighting} add antidote/fake data to the training dataset following data poisoning attacks, while 
in ~\cite{ekstrand2018all} they use re-sampling to adjust the proportion of users in groups.

The benefit of pre-processing methods is that they only change the input data and thus existing models can be applied to the adjusted dataset, providing significant flexibility. However, while mitigating bias from the dataset, relevant information to the downstream task might also be removed. This will result in uncertainty in the tradeoff between utility and fairness.

\subsubsection{In-processing Methods}

Various techniques can guide the training process of the model towards higher utility and better fairness, we introduce the two major directions.

\noindent\textbf{Optimization with Regularization and Constraint:} 
Regularization-based methods integrate fairness with utility in the optimization objective by introducing a fairness regularizer~\cite{tang2023fairness, kamishima2013efficiency, wan2020addressing}. The new overall loss term is formulated with a Lagrange multiplier as $\mathcal{L}=\mathcal{L}_{\text{rec}}+\lambda\mathcal{L}_{\text{fairness}}$,  where $\mathcal{L}_{\text{rec}}$ is the general recommendation objective, $\mathcal{L}_{\text{fairness}}$ is the fair regularization, and $\lambda$ is the coefficient to balance two goals. Some works add an independence regularizer to encourage the recommendation to be independent of the sensitive features. Such independence terms include mean matching~\cite{kamishima2013efficiency}, distribution matching~\cite{kamishima2018recommendation}, and mutual information~\cite{kamishima2018recommendation}. Error correlation loss~\cite{wan2020addressing}  is designed to regularize the correlation
between prediction errors and the distribution of market segments. Four unfairness metrics~\cite{yao2017beyond} are proposed as the regularizer, measuring the discrepancy between the prediction behavior of the disadvantaged and advantaged groups. \cite{yang2023towards} advanced a similar unfairness loss based on Distributionally Robust Optimization(DRO) technique. A tensor-based fairness-aware RS (FATR)~\cite{zhu2018fairness} is proposed that adds an orthogonality term between the representations of users/items and the corresponding vectors of sensitive features. Counterfactual graphs are generated and utilized in the regularization~\cite{chen2024fairgap}. \cite{chen2023improving} add regularization based on data augmentation via generating `fake' interaction data. \cite{ying2023camus} add regularization by conducting data augmentation for minority group which utilizes interactions of mainstream users.

The main idea of constrained optimization~\cite{singh2018fairness,singh2019policy} is similar to regularization-based methods where fair constraints are included during optimization. However, it has a different form as \textit{${\text{minimize}}\quad \mathcal{L}_{\text{rec}}\quad\text{s.t.} \text{~~Fairness constraints}$}. The difference between optimization with regularization and constraints is that the former tolerates unfairness where the solution might fall into unfair regions but the latter disallows unfairness enforced by the fairness constraints.

\noindent\textbf{Adversarial learning}: The main idea is to learn fair representations irrelevant to sensitive features~\cite{wu2021learning, wang2022improving}. To achieve this, a discriminator to predict the sensitive label and a generator to generate fair representations are trained. During adversarial learning, the discriminator gains the ability to predict the sensitive label while the generator is optimized to fool the discriminator by generating fair representations so that the discriminator cannot determine whether the representation contains sensitive features. By playing the min-max game between the discriminator and generator, sensitive information will be removed from the final learned representations and only relevant information for the downstream task is maintained. There have been many works in this direction~\cite{wu2021learning, wang2022improving,dai2021say, grari2023adversarial,chen2023fmmrec,hua2023up5,hu2023automatic}, primarily following the aforementioned setup.

The in-processing methods allow more control in utility-fairness trade-off during training. However, they might be designed for specific models and cannot be generalized to other models.

\subsubsection{Post-processing Methods}

Re-ranking methods are widely used, as post-processing approaches, to adjust the recommendations generated by the model to promote fairness~\cite{wang2023survey,steck2018calibrated,xiao2017fairness,li2021user}. According to the granularity of adjustment at each time, there are three different re-ranking types~\cite{wang2023survey}: Slot-wise re-ranking~\cite{steck2018calibrated,karako2018using}, User-wise re-ranking~\cite{xiao2017fairness,biega2018equity}, and Group-wise re-ranking~\cite{li2021user,fu2020fairness}.

\noindent\textbf{Slot-wise re-ranking:} These methods add items from candidate list sequentially to recommendation list (i.e., item by item) by following rules considering relevance and fairness simultaneously. The greedy algorithm~\cite{steck2018calibrated,karako2018using} to select the item with the maximum marginal gain is as follows:
\begin{equation*}
\small 
    i^* = argmax_{i\in \mathcal{C}\setminus \mathcal{S}} \lambda f_\text{rec}(u, \mathcal{S} \cup \{i\})+(1-\lambda)f_\text{fair}(u, \mathcal{S} \cup \{i\}),
\end{equation*}
\vspace{-0.5ex}
where $\mathcal{C}$ is the candidate item set with high relevance scores calculated by the trained RS, and $\mathcal{S}$ is the current generated recommendation list. Function $f_\text{rec}$ measures utility, and $f_\text{fair}$ measures fairness. The item which will bring the largest contribution to the existing recommendation list will be added to the list from $\mathcal{C}$, leading to an updated list $\mathcal{S} = \mathcal{S}\cup\{i^*\}$. The items will be iteratively added until the recommendation list reaches the required length.

\noindent\textbf{User-wise re-ranking:} These methods generate the whole recommendation list for a user at once. A popular strategy is integer programming~\cite{wolsey2020integer,conforti2014integer}. The main idea is to treat decisions as variables (e.g., whether an item is in the recommendation list~\cite{xiao2017fairness} or is in a particular position~\cite{biega2018equity}) and transform the re-ranking problem into an integer programming problem with fairness constraints.

\noindent\textbf{Group-wise re-ranking:} Re-ranking at a group level considers several users together rather than adjusting the list for a single user. Similar to user-wise re-ranking, integer programming can also be leveraged. Li et al. ~\cite{li2021user} use a binary matrix $\mathbf{W}\in\mathbb{R}^{|\mathcal{U}|\times |\mathcal{I}|}$ to represent whether an item will be recommended in the top $K$ list for each user. They solve the following optimization problem:
\begin{align*}
\small 
\underset{\mathbf{W}}{\text{max}} \sum_{i=1}^{|\mathcal{U}|}\sum_{j=1}^{|\mathcal{C}|} \mathbf{W}_{ij}\mathbf{R}_{ij}, \quad \text{s.t. GUF}(Z_1, Z_2, \mathbf{W})<\epsilon,  \sum_{j=1}^{|\mathcal{C}|}\mathbf{W}_{ij}=K, \mathbf{W}_{ij}\in \{0,1\},
\end{align*}
where GUF is the user fairness constraint between two groups and $\mathbf{R}_{i j}$ is the relevance score which predicts the user $i$'s preference towards item $j$. The optimization problem aims to maximize the relevance score while subjecting to fairness constraint. Another work~\cite{fu2020fairness} uses a similar strategy in explainable recommendation with knowledge graphs and adds another fairness constraint for explanation fairness, which is an area with growing attention~\cite{grabowicz2022marrying,zhao2022fairness}. 

The post-processing methods provide model-agnostic flexibility. However, the improvement is restricted by the results from base models, leading to a suboptimal solution. In other words, if the base model generates extremely biased results, the adjustment based on it might be limited.

\section{Diversity in Recommender Systems}
\label{sec-diversity}

As another essential beyond-utility perspective, diversity is comprehensively discussed here from three dimensions: (1) the source of diversity, (2) the measurements to quantify diversity, and (3) the methods to promote diversity. Figure~\ref{fig:measurements-and-methods}(b) summarizes the diversity measurements and methods. Various other concepts such as \textit{Serendipity} and \textit{Novelty} are closely related to diversity but slightly different~\cite{aggarwal2016evaluating}. We refer readers to~\cite{kaminskas2016diversity} if interested in a comprehensive discussion on these concepts.

\subsection{Source of Item Diversity}
Diversity is generally discussed from the item side~\cite{ziegler2005improving, kelly2006enhancing, vargas2011rank, ribeiro2012pareto, shani2011evaluating}. In most papers, it is defined based on the redundancy or similarity among the recommended items, where the detailed difference might come from the categories~\cite{vargas2014coverage, chen2007preference} or the distance in item embedding space~\cite{vargas2011intent, shi2012adaptive}. Items are naturally different from each other, and the level of diversity is different across items. We will discuss item diversity from three aspects: (1) When category information is available, the items belonging to the same category share higher similarity than items from different categories. For example, two movies both from romance will generally be more similar compared to two motives from romance and horror genre respectively. (2) Within the same category, items might still have different features. Take movies as an example, two movies are both categorized as romance, but they have different stories, topics, budgets, language, etc. (3) Besides the intrinsic features, item diversity is implicitly revealed from users' history interactions. According to the collaborative filtering effect, similar users prefer similar items. Therefore, user interactions can be utilized as diversity indicator. In summary, item diversity might come from categories, at a high level, or from item features, at a detailed level. Item features can be explicit (i.e., intrinsic item features) or implicit (i.e., from learned embeddings considering both item features and other users' history interactions).

\subsection{Measurements to Quantify Item Diversity}
\label{sec-diversity_metrics}
Diversity can be categorized into individual diversity (i.e., individual-level) which focuses on a single recommendation list and is relevant to each user's satisfaction individually~\cite{kelly2006enhancing,shi2012adaptive,ribeiro2012pareto}, and aggregate diversity (i.e., system-level) which aims to capture the diversity across all recommendations and is relevant to the fairness of providers~\cite{adomavicius2011improving,fleder2007recommender,shani2011evaluating,xu2023p}. 

\subsubsection{Individual Diversity} Individual diversity can be further split into distance-based and category-based. They rely on the distance between different item representations or genre information.

\noindent\textbf{Distance-based diversity:} The most widely used definition is called \textit{Intra-List Diversity (ILD)} which measures the pairwise item diversity within one recommendation list. 

\begin{definition}
\small 
    \textbf{Intra-List Diversity (ILD)} is formally defined as follows where the distance function $d$ measures the dissimilarity/distance between items:
    \begin{equation}
    ILD(\mathcal{R}_u) = \frac{1}{|\mathcal{R}_u|(|\mathcal{R}_u|-1)}\sum_{i \in \mathcal{R}_u}\sum_{j \in \mathcal{R}_u \setminus {i}}d(i, j).
    \label{eq.ild}
\end{equation}

\end{definition}
The variants differs in the way of obtaining item embeddings and the specific distance function. Items can be represented by content descriptors~\cite{ziegler2005improving}, rating scores~\cite{kelly2006enhancing, vargas2011rank}, latent item representations from matrix factorization~\cite{vargas2011intent, shi2012adaptive}, etc. 
For distance, various functions have been applied, e.g.,
Hamming distance~\cite{kelly2006enhancing}, Gower diversity~\cite{haritsa2009kndn}, complement of cosine similarity~\cite{ribeiro2012pareto}, Jaccard similarity~\cite{vargas2011rank}, or Pearson correlation~\cite{vargas2011rank}. A larger \textit{ILD} indicates a higher level of diversity.

\noindent\textbf{Category-based diversity: } Distance-based diversity is criticized in \cite{vargas2014coverage} due to the failure of ensuring the consistency between the diversity value and users' experience. 
They~\cite{vargas2014coverage} leverage category/genre information to capture item diversity, which better corresponds to users' perceptions. They propose a novel binomial framework to capture genre-based diversity ~\cite{vargas2014coverage} which considers three perspectives simultaneously: coverage, redundancy, and size-awareness. When genre information is available, they compute the diversity score as $\text{BinomDiv}(\mathcal{R}_u) = \text{Coverage}(\mathcal{R}_u) * \text{NonRed}(\mathcal{R}_u)$ where coverage mainly captures how many different genres are presented in the recommendation and non-redundancy (i.e., NonRed) encourages the genre uniqueness in the recommendation. When explicit category information is unavailable, Chen et. al~\cite{chen2007preference} propose to group items into categories based on their attributes. Thereafter, diversity is defined as the category dissimilarity.

\subsubsection{Aggregate Diversity}
Unlike individual diversity which corresponds to a single recommendation list, aggregate diversity considers all recommendations in an aggregated way. If a RS always recommends popular items rather than a diversified list, from a global view, the aggregate diversity would be poor. Therefore, it reflects the system's ability to recommend less popular or hard-to-find items and thus is related to exposure fairness~\cite{singh2018fairness,borges2021mitigating,wu2021tfrom}. The most intuitive definition is to count the number of total diverse items being recommended~\cite{adomavicius2011improving}.
\begin{definition}
    \textbf{Aggregate Diversity (Count)} is defined as the length of the recommendation set of all users: $Aggdiv=|\bigcup_{u\in \mathcal{U}}\mathcal{R}_u|$. 
\end{definition}
This measurement focuses on whether the item is recommended but ignores how many users this item is recommended to, which motivates other work to investigate whether such recommendation is evenly distributed when taking the detailed user number into account. 
For instance, aggregate diversity has been formulated with Shannon entropy (H)~\cite{shani2011evaluating} and Gini index (Gini)~\cite{fleder2007recommender,shani2011evaluating}. 

\begin{definition}
    \textbf{Aggregate Diversity (Shannon Entropy)} is defined as:
\begin{equation}
    H=-\sum_{i \in \mathcal{I}}p(i)log_{2}p(i), \quad p(i)=\frac{|\{u \in \mathcal{U}|i\in \mathcal{R}_u\}|}{\sum_{j \in \mathcal{I}}|\{u \in \mathcal{U}|j\in \mathcal{R}_u\}|},
    \label{eq.entropy}
\end{equation}
where $p(i)$ measures the probability of item $i$ being in the recommendation list for all users.
\end{definition} 

\begin{definition}
\textbf{Aggregate Diversity  (Gini Index)} is defined as:
    \begin{equation}
    \small 
    Gini=\frac{1}{|\mathcal{I}|-1}\sum_{k=1}^{|\mathcal{I}|}{(2k-|\mathcal{I}|-1)}p(i_k),
    \label{eq.gini}
\end{equation}
where $p$ shares the same meaning as in the Shannon entropy formula.
\end{definition}

\begin{figure*}[t]
    \centering
    \includegraphics[width=0.8\textwidth]{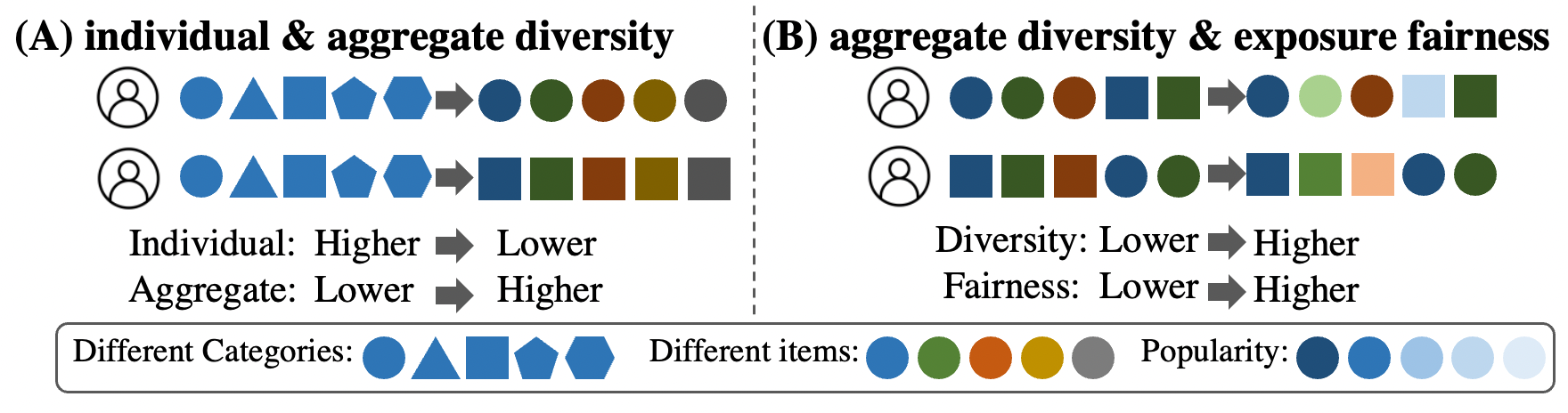}
    \vskip -1.5ex
    \caption{(A) A toy example to show the difference between individual and aggregate diversity; (B) A toy example to show the connection between aggregate diversity and exposure fairness. Different shapes correspond to different categories, colors correspond to different items, and levels of the darkness of the same color correspond to different levels of popularity (the darker color indicates the more popular item).}
    \label{fig:diversity_illustration}
    \vspace{-3ex}
\end{figure*}

\subsubsection{Discussion}
Individual and aggregate diversity measure diversity from two distinct levels: individual and system, respectively. High individual diversity does not imply high aggregate diversity and vice versa. For example, in Figure~\ref{fig:diversity_illustration}(A), the recommendation could be of high individual diversity for each user but of low aggregate diversity from a system level, or the recommendation could be of low individual diversity for each user but provide high aggregate diversity.

\subsection{Methods to Promote Diversity}
Various methods have been proposed to promote diversity. They can be categorized into four types~\cite{castells2021novelty}: re-ranking-based, learning to rank, cluster-based, and fusion-based.

\begin{center}
\vspace{0.5ex}
\begin{minipage}{.68\linewidth}
\begin{algorithm}[H]
 \DontPrintSemicolon
 \footnotesize
 \KwIn{Recommendation number top $K$, coefficient $\lambda$}
 \KwRet{Recommendation list $\mathcal{R}_u$}
 
    Init recommendation list $\mathcal{R}_u=\{\}$
    
    \For{$|\mathcal{R}_u| < K$}
    { 
    $ i^* = argmax_{i \in \mathcal{I}\setminus \mathcal{R}_u}s(\mathcal{R}_u \cup \{i\}, \lambda)$

    $\mathcal{R}_u=\mathcal{R}_u \cup \{i^*\}$

    }
    \KwRet{$\mathcal{R}_u$};
\caption{\small Greedy Algorithm for Diversity Enhancement}
 \label{alg.greedy}
\end{algorithm}
\end{minipage}
\vspace{1ex}
\end{center}

\noindent\textbf{(1) Re-ranking} 
methods aim to adjust the ranking from existing RS by combining the diversity constraints to improve individual or aggregate diversity. The traditional rank score is solely based on relevance, while the new one~\cite{ziegler2005improving, kelly2006enhancing} is composed of relevance and diversity in the form of:
\begin{equation}
\small 
s(\mathcal{R}_u, \lambda)=\frac{1-\lambda}{|\mathcal{R}_u|}\sum_{i\in \mathcal{R}_u}f_{\text{rec}}(i)+\lambda f_{\text{div}}(\mathcal{R}_u),
    \label{eq.rerank}
\end{equation}
where $f_{\text{rec}}(i)$ denotes the relevance score of item $i$, $f_\text{div}(R)$ is the diversity score of the recommendation list $\mathcal{R}_u$, and $\lambda$ is the coefficient to trade-off the utility and diversity goals. As discussed in Section~\ref{sec-diversity_metrics}, various diversity metrics can be adopted for computing $f_\text{div}(\mathcal{R}_u)$. After having the new score, a greedy algorithm~\cite{carbonell1998use} called Maximum Marginal Relevance (MMR) is performed iteratively to select the item with the maximum score until reaching the expected recommendation length as shown in Algorithm~\ref{alg.greedy}. Many works follow this greedy framework~\cite{smyth2001similarity, ziegler2005improving, kelly2006enhancing, vargas2014coverage, santos2010exploiting}.

Beyond the greedy framework, other re-ranking algorithms mainly rely on solving constraint optimization problems to find the optimal list. \cite{zhang2008avoiding} transforms the problem into a series of objective functions with different utility-diversity constraints (e.g., maximizing the relevance under the constraint that diversity is larger than a diversity tolerance, maximizing the diversity when the relevance is larger than a matching tolerance). The benefits of these post-processing methods are that they are model-agnostic and time-efficient (i.e., avoid the computation of re-training). However, the improvement might be restricted by the initial recommendation list.

\noindent\textbf{(2) Learning to rank} methods, are in-processing methods that enforce adjustment during training process. Typical RS are trained with utility loss functions such as Bayesian Personalised Ranking (BPR) loss~\cite{rendle2012bpr} or rank-based loss which aims to improve utility performance, several works add diversity objectives to the existing one so that during the training, the model is trained towards higher utility and to avoid monotony simultaneously. \cite{wasilewski2016incorporating} continues and explores the work in \cite{hurley2013personalised}, which incorporates diversity criteria in regularization terms, by proposing several regularizations on learned item embeddings to incorporate diversity. The regularizations are as follows: 
\begin{equation}
\small 
    reg(\mathbf{E}_U, \mathbf{E}_I) = \sum_{i, j}d(i, j)\|\mathbf{e}_I^i-\mathbf{e}_I^j\|^2, \quad 
    = \sum_{u, i, j}d(i, j)({\mathbf{e}_U^u}^\top(\mathbf{e}_I^i-\mathbf{e}_I^j))^2, \quad 
    \text{or } = \sum_{u, i, j}d(i, j){\mathbf{e}_I^i}^\top \mathbf{e}_I^j,
\end{equation}
where $\mathbf{E}_U$ is the learned user representation, $\mathbf{E}_I$ is the learned item representation, and $d(i, j)$ is a pre-defined distance between item $i$ and item $j$. Generally, when the distance $d(i, j)$ is large, the representations between these two items will be trained to be closer due to the regularization. Therefore, more diverse items will be recommended. \cite{li2020cascading} takes genre diversity into account in the reward model for a multi-armed bandit recommendation.
More recently, diversity-aware deep ranking network~\cite{wang2023diversity} is proposed to generate accurate and diversified recommendation list during ranking phase.
Compared with the post-processing methods, this line of research involves re-training, which might be less efficient. The advantage is that it can ensure diversity in the recommended lists, owing to incorporating diversity considerations during training. 

\noindent\textbf{(3) Cluster-based} methods leverage the principle that similar items will be grouped into the same cluster, and therefore to promote diversity, items from different clusters rather than a single cluster should be recommended. Following this idea, different methods are proposed to select items from various clusters~\cite{zhang2009novel, aytekin2014clustering, berbague2021overlapping, li2012multidimensional}. For example, \cite{zhang2009novel} clusters items based on user profiles and recommend items that match these individual clusters rather than the whole user profile.
\cite{aytekin2014clustering} 
proposes ClusDiv which assigns how many items (i.e., weight) each cluster should be recommended for each user. It initially assigns the weights based on the recommendation list from traditional RS and then adjusts the weights by iteratively decreasing the number of items within the category that is larger than a threshold and increasing the number of items in the least recommended category. After having the adjusted weights, items based on this assignment will be selected for recommendation. For comparison, \cite{zhang2009novel} performs clustering based on local information according to user's tastes while \cite{aytekin2014clustering} is based on global information.

\noindent\textbf{(4) Fusion-based} methods aggregate results from different RS. While a single RS might provide recommendations with high utility performance but low diversity, different RS will provide high-quality recommendations while obtained recommendations are dissimilar from each other. Therefore, model fusion can be leveraged to obtain a result with both high utility and diversity. \cite{ribeiro2012pareto} fuses the rating scores to generate new aggregated scores. It proposes a multi-objective framework that considers utility, diversity, and novelty simultaneously. It first adopts multiple existing RS to generate the predicted ratings for user-item pairs and fuses the predictions by a weighted summation $\hat{r_{ui}}=\sum_{t=1}^{T}w_t r_{ui,t}$, where $T$ denotes different RS, $r_{ui,t}$ is the rating score estimated from the $t$-th RS, and the weights $w_t$ is the strength for considering the corresponding model during aggregation, which are learnable with a strength Pareto evolutionary algorithm~\cite{zitzler1999multiobjective,zitzler2001spea2}.
The aggregated ratings $\hat{r_{ui}}$ are then used for generating the recommendation list.
\section{Fairness and Diversity}
\label{sec-fd}
As different beyond-utility perspectives, fairness and diversity are generally investigated separately. However, there are various connections that will be discussed in Section.\ref{subsec.connections}. Based on the connections and differences, we will first summarize diversity from user perspective and then discuss the fairness works for user diversity and item diversity.

\subsection{Connections and Differences}
\label{subsec.connections}
\begin{figure*}
    \centering
    \includegraphics[width=0.75\textwidth]{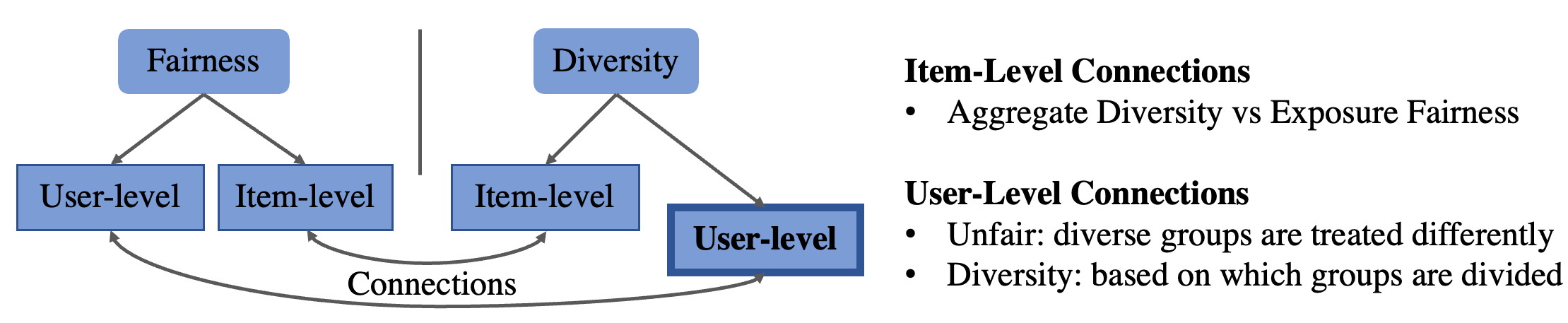}
    \vskip -1.5ex
    \caption{Fairness and Diversity: in the context of recommender systems, they are commonly investigated separately. However, the connections at item and user level highlight the significance of intersections.}
    \label{fig:fairness_and_diversity}
    \vskip -2.0ex
\end{figure*}

With the rapid development of RS, utility performance (e.g., accuracy) is no longer the sole golden standard for determining the quality of the system. There have been various extra considerations which are closely related to users' satisfaction. Fairness and diversity are two such beyond-utility perspectives with each highlighting different aspects. From an ethical perspective, fairness and diversity commonly appear in the same sentence as they are similar in their meaning. Specifically, bias exists when diverse groups are treated dissimilarly and fairness requires these diverse groups to receive similar treatments. From a restricted standpoint, such diversity refers to sensitive features (e.g., race, gender, etc.) which is typically discussed in the fairness field. However, there are also fairness works beyond discussing such sensitive features. They can be summarized in a more comprehensive view. From a more broad standpoint, there are various diversities based on which fairness is proposed. As fairness is generally discussed from both the user and item sides, we naturally discuss connections from the item level and user level as illustrated in Figure~\ref{fig:fairness_and_diversity}.

\noindent\textbf{Item-level}: as mentioned in Section~\ref{sec-diversity}, aggregate diversity is related to exposure fairness. When a RS tends to always recommend popular items which shows a lack of exposure fairness, the aggregate diversity will be low since there will be a large overlapping of popular items. When improving exposure fairness by recommending more unpopular items, the chances that the same unpopular item being recommended to different users will be low, therefore, the aggregate diversity will probably increase. A toy example is shown in Figure~\ref{fig:diversity_illustration}(B).

\noindent\textbf{User-level}: user-level connection is less intuitive than item-level. One of the most important reasons is that typical diversity is discussed from the item perspective. As fairness is discussed at both item and user levels, we first summarize diversity from a user perspective, namely, user diversity, which is rarely discussed in the previous literature. User diversity includes explicit/implicit features, historical preferences (proportionality), fairness requirements, and multiple interests (general) and will be summarized in Section~\ref{subsec.user_diversity}. From the first three perspectives, there are corresponding fairness studies related to that type of user diversity, indicating their strong connections. The last user diversity opens up future direction for fairness in terms of multiple interests.

\subsection{User Diversity}
\label{subsec.user_diversity}
While diversity is typically discussed within the context of items, diversity also exists from the aspect of the user side. Users have diverse preferences and interests, which are reflected in their history interactions. In this section, we summarize the user-level diversity from \textit{explicit/implicit features, historical preferences (proportionality), fairness requirements, and multiple interests (general)}. While historical preferences and multiple interests are all related to user preference, the former focus on genre proportionality and the latter is more general.

\noindent\textbf{Explicit/Implicit Features:}
Users have diverse properties in terms of their inherent features (i.e., explicit features) and behavior features extracted from their interactions (i.e., implicit features). Inherent features include age, gender, race, etc. Behavior features include the interaction number, the average price from interacted items, and so on. These features are sometimes treated as sensitive features from a fairness perspective~\cite{ferraro2021break,islam2021debiasing,wu2021learning,li2021user,wu2022big}

\noindent\textbf{Historical Preferences (Proportionality):}
Users have diverse preferences which are reflected in their history interactions. Specifically, it is reflected in the proportion of interacted genres/categories in the history logs. For example, two users who watched both romance movies and action movies will share similar interests. However, their focus might be different as one user watched 70 romance movies and 30 action movies while another user watched 50 romance and 50 action movies. It is reasonable to expect the recommender system to provide a personalized recommendation list that reflects such differences. This important property is known as calibration~\cite{steck2018calibrated}, which is proposed to avoid the issue of utility-oriented systems where the user's lesser interest gets crowded out by the main interest. The idea of proportionality was first proposed in ~\cite{dang2012diversity}.

\noindent\textbf{Fairness Requirements:}
Although it is obvious to perceive the users' diversified needs that results in the personalization requirements on recommender system, the other key aspect of this survey, fairness, receives different attitudes from users. First, users would have different tolerance levels of fairness. For example, in the microlending platform~\cite{liu2019personalized}, lenders’ tolerance of fair consideration related to regions varies greatly. Some lenders prefer offering loans to certain regions like their home countries, while others may be open to diverse
regions. Secondly, users might treat different features as sensitive features. For instance, some users treat gender as a sensitive feature since they do not want the recommendations to be influenced by this feature, while others may care more about the age feature than the gender feature~\cite{li2021towards}.

\noindent\textbf{Multiple Interests:}
Unlike focusing on the proportionality of interacted genres, interests capture more general user preference towards certain items. It provides a high-level depiction of the user. Traditional RS learn a single embedding to represent the user. Recently, researchers have proposed that this will lose information during aggregation and cannot fully depict users' diverse interests. Therefore, a line of research focuses on using multiple embeddings to represent users' diverse interests~\cite{cen2020controllable,li2019multi}. Additionally, a similar topic is called disentangled learning~\cite{ma2019disentangled, wang2020disentangled}, where the single embedding is disentangled to multiple sub-embeddings such that these latent sub-embeddings would reflect different intentions towards various items. While users have multiple interests/intentions during decision-making, different users have diversified interests. 

\subsection{Fairness for User Diversity}
We discuss fairness works for proposed diversity, which are summarized in Table~\ref{tab:fairness_user_diversity} and  Fig.~\ref{fig:fairness_and_diversity_visualization}. 

\begin{table*}[]
    \centering
    \setlength\tabcolsep{3pt}
    \tiny
    \caption{\textcolor{blue}{A summary table of fairness for user diversity. Papers and available codes can be accessed from}\href{https://github.com/YuyingZhao/Awesome-Fairness-and-Diversity-Papers-in-Recommender-Systems}{\textcolor{blue}{link}}.}
    \label{tab:fairness_user_diversity}
    \vskip -2.5ex
    \begin{tabular}{|p{3cm}|c|c|c|c|c|c|}
        \hline
        Fairness (Diversity) & \multicolumn{2}{c|}{Work} & Intervention & Approach & Keywords & Code \\
        \hline
        \multirow{16}{3cm}{General Fairness \\ (Explict/ Implict Features)}  &  \multirow{4}{*}{Gender} & \cite{deldjoo2021flexible} & - & - & Fairness Evaluation & \xmark \\
          &  & \cite{wu2021fairness} & In-processing & Adversarial learning & News Recommendation & \cmark \\
          &  & \cite{xia2019we} & In-processing & Constraint-based & Reciprocal & \xmark\\
          &  & \cite{burke2018balanced} & In-processing & Regularization-based & Multi-side fairness & \xmark \\
          \cline{3-7}
          &  \multirow{4}{*}{Race} & ALG~\cite{gorantla2021problem} & Post-processing & Re-ranking & Theoretical & \cmark \\
          &  & \cite{ghosh2021fair} & - & - & Uncertain Inference & \cmark\\
          &  & MSRec~\cite{zheng2018fairness} & Post-processing & Re-ranking & Dating & \xmark\\
          &  & FATR~\cite{zhu2018fairness} & In-processing & Constraint-based & Tensor-based & \cmark\\
          \cline{3-7}
          &  \multirow{2}{*}{Age} & \cite{deldjoo2021flexible} & - & - & Fairness Evaluation & \xmark\\
          &  & PSL~\cite{farnadi2018fairness} & In-processing & Logical rules & Hybrid & \xmark\\
          \cline{3-7}
          &  \multirow{4}{*}{Behavior} & \cite{li2021user} & Post-processing & Constraint-based & User activeness & \cmark\\
          &  & \cite{hao2021pareto} & In-processing & Constraint-based & Pareto Optimality & \xmark \\
          &  & \cite{fu2020fairness} & Post-processing & Constraint-based & Explainable & \cmark\\
          &  & PFGR~\cite{xiao2020enhanced} & In-processing & Constraint-based & Group recommendation & \xmark \\
        \hline
        
        \multirow{5}{3cm}{Calibration-based Fairness \\ (Historical Preferences - proportionality)}  &  \multicolumn{2}{c|}{\cite{steck2018calibrated}} & Post-processing & Re-ranking & Calibrated & \xmark\\
        &   \multicolumn{2}{c|}{\cite{abdollahpouri2019unfairness}} & - & - & Popularity bias & \xmark \\
        &   \multicolumn{2}{c|}{\cite{abdollahpouri2021user}} & Post-processing & Re-ranking & Popularity bias & \xmark \\
        &  \multicolumn{2}{c|}{\cite{da2021exploiting}} & Post-processing & Re-ranking & Multiple fairness metrics & \cmark \\
        &  \multicolumn{2}{c|}{\cite{zhao2021rabbit}} & Post-processing & Re-ranking & Taste distortion & \xmark\\
        
        \hline
        \multirow{6}{3cm}{Personalized Fairness \\ (Fairness Requirements)}  &  \multicolumn{2}{c|}{FAR/PFAR~\cite{liu2019personalized}} & Post-processing & Re-ranking & Diversity tolerance & \xmark \\
        &   \multicolumn{2}{c|}{OFAiR~\cite{sonboli2020opportunistic}} & Post-processing & Re-ranking & Multi-aspect fairness & \xmark\\
        &   \multicolumn{2}{c|}{\cite{li2021towards}} & In-processing & Adversarial learning& Causal notion & \cmark\\
        &   \multicolumn{2}{c|}{PFRec~\cite{wu2022selective}} & Post-processing & Adversarial training & Prompt-based & \cmark\\
        &  \multicolumn{2}{c|}{UCRS~\cite{wang2022user}} & In-processing & Counterfactual inference & Filter Bubbles & \cmark\\
        &   \multicolumn{2}{c|}{CUFRL~\cite{cui2023controllable}} & In-processing & Regularization-based & Fair representation & \xmark\\
        
        \hline
        \multirow{8}{3cm}{Multi-Interest Fairness\footnotemark \\ (Multiple Interests - general)}  &  \multicolumn{2}{c|}{MIND~\cite{li2019multi}} & - & - & Dynamic routing & \cmark\\
        &   \multicolumn{2}{c|}{ComiRec~\cite{cen2020controllable}} & - & - & Self-attention & \cmark \\
        &   \multicolumn{2}{c|}{Re4~\cite{zhang2022re4}} & - & - & Backward flow & \cmark\\
        &  \multicolumn{2}{c|}{PinnerSage~\cite{pal2020pinnersage}} & - & - & Cluster-based & \xmark\\
        &   \multicolumn{2}{c|}{MIP~\cite{shi2022every}} & - & - & Time-aware & \cmark\\
        &   \multicolumn{2}{c|}{MacridVAE~\cite{ma2019learning}} & - & - & VAE & \cmark \\
        &   \multicolumn{2}{c|}{DisenGCN~\cite{ma2019disentangled}} & - & - & Neighborhood routing &\cmark \\
        &   \multicolumn{2}{c|}{DGCF~\cite{wang2020disentangled}} & - & - & Collaborative filtering & \cmark\\
        \hline
    \end{tabular}
    \vskip -5ex
\end{table*}
\footnotetext{To the best of our knowledge, there are no related multi-interest fairness works and we review works related to multi-interest. Due to the same reason, while summarizing these works, we will not categorize them in the intervention and approach columns designed for fairness works.}

\subsubsection{General Fairness}
Most works related to user fairness fall into this category where groups are divided by explicit sensitive features (e.g., gender, race, religion)~\cite{ferraro2021break,islam2021debiasing,wu2021learning} or implicit sensitive features (e.g., degree)~\cite{li2021user,wu2022big}. The goal is to ensure that diverse groups would receive similar treatments. To achieve this, methods discussed in Section~\ref{sec.debias}, such as regularization-based optimization and adversarial learning, can be applied. Specifically, these works focus on the explicit sensitive features which annot be easily changed by users, including gender~\cite{deldjoo2021flexible,wu2021fairness,xia2019we, burke2018balanced}, race~\cite{gorantla2021problem,ghosh2021fair,zheng2018fairness,zhu2018fairness}, age~\cite{deldjoo2021flexible, 
farnadi2018fairness}. Other works focus on implicit features extracted from user behaviors (e.g., interaction number, price of purchased items)~\cite{li2021user, hao2021pareto,fu2020fairness,xiao2020enhanced}.

\subsubsection{Calibration-based Fairness}
Calibration in classification requires that the predicted class distribution aligns with the input data's actual distribution. Similarly, calibration in recommendations aims to ensure that the genre distribution in the recommended list aligns with that in users' historical interactions, thereby more accurately reflecting users' interests. For instance, in movie recommendations encompassing various movie categories like action and romance, if a user watches $90\%$ action movies and $10\%$ romantic movies, the recommendation list should reflect this property rather than exclusively recommending all action movies. Recommending items without following users' history preference would lead to unfairness in terms of preserving preference proportionality.
To better preserve preference proportionality, Steck~\cite{steck2018calibrated} propose calibrated recommendation with post-processing methods to adjust the genre distribution with maximum marginal relevance~\cite{carbonell1998use}. The main idea is to first obtain recommendations with baseline recommender systems and then update the list to fit the distribution in historical interactions. As demonstrated in Eq.~\ref{eq.cali}, the adjusted recommendation list $\mathcal{R}_u^*$ considers two aspects (1) utility, indicated by the relevance score $f_{\text{rec}}(\mathcal{R}_u)$, (2) fairness, indicated by the distribution divergence where $p$ denotes the genre distribution in users' historical interactions and $q$ represents the distribution in the recommendation and $C_{KL}$ measures the Kullback–Leibler(KL)-divergence between these two distributions.
\begin{equation}
    \mathcal{R}_u^* = argmax_{\mathcal{R}_u, |\mathcal{R}_u|=K} (1-\lambda)f_{\text{rec}}(\mathcal{R}_u)-\lambda C_{KL}(p, q(\mathcal{R}_u))
    \label{eq.cali}
\end{equation}
Minimizing the divergence allows the recommendation to better align with user interests reflected in interactions, with $\lambda$ controlling the trade-off between utility and fairness. Inspired by Steck's work, Abdollahpouri et al. ~\cite{abdollahpouri2019unfairness} investigate unfairness of popularity bias from the user perspective. They discover that the property of consistent genre distribution in recommendation and interactions is maintained at varying levels for users with diverse interests in popular items. Users are categorized into three groups based on the ratio of popular items in their historical interactions: Niche Users (those with the least interest in popular items), Blockbuster-focused Users (those with the highest interest in popular items), and Diverse Users (all others). They find that Niche Users suffer more from popularity bias. Because although these users are interested in unpopular/long-tail items, the recommendation still concentrates on popular items. This finding suggests that the preferences of Niche Users are not as well-served as those of Blockbuster-focused Users. After observing this phenomenon, in a follow-up work~\cite{abdollahpouri2021user}, Abdollahpouri et al. design a metric called User Popularity Deviation (UPD) to quantify the popularity bias from a user-centered view and propose a re-ranking method called Calibrated Popularity which is similar to Eq.~\ref{eq.cali} with a different fairness term based on Jensen–Shannon divergence. Silva et al.~\cite{da2021exploiting} conduct a comprehensive experiment to investigate calibration in fair recommendations. They evaluate six recommender algorithms applied in the movie domain and analyze variations of three fairness measures, including three distribution metrics: Kullback–Leibler, Hellinger, and Pearson chi-square~\cite{cha2007comprehensive}. Previous works generally assume static user interest. However, Zhao et al.~\cite{zhao2021rabbit} emphasize that interest would evolve over time. They extend the static setting by predicting the future genre distribution that matches the user interest and conducting the calibration based on the predicted distribution rather than the one extracted from historical interactions.

\vspace{-1ex}
\subsubsection{Personalized Fairness}
Users have diverse fairness demands. They have different levels of demands/tolerance on fairness consideration. For example, some users put more emphasis on fairness than others. Additionally, individuals also differ in which features should be considered sensitive. For instance, some people think age is a regular feature that helps recommend popular songs among their peers and some consider it as sensitive as they would like to follow the trend and not limited by age. Therefore, this suggests fairness may need to be personalized to individuals. Liu et al.~\cite{liu2019personalized} investigate personalized fairness in microlending, where recommender systems are designed to recommend loans to lenders. In the design, besides the fairness need that borrowers from diverse demographic groups should have a fair chance of being recommended, they also consider personalized fairness on the lender side. Lenders' receptivity to the diversification of recommended loans varies greatly. Some lenders prefer certain regions while others are open to diverse areas. They use the information entropy~\cite{shannon2001mathematical} to identify the lender diversity tolerance and assign it as the weight of fairness term in the re-ranking process so that different lenders show personalized focus on the fairness aspects. Sonboli et al.~\cite{sonboli2020opportunistic} follow their work and extend it into the scenario with multiple sensitive features. They propose a more fine-grained personalization according to each feature. For different protected features, users have different tolerance which is also measured by information entropy. Li et al.~\cite{li2021towards} study personalized fairness based on the causal notion. Users are allowed to specify sensitive features. Feature-independent user embeddings are generated so that the recommendation outcomes maintain the same in the counterfactual world where the other features are unchanged except for specified sensitive features. They design an adversarial learning approach to remove sensitive information while keeping relevant information for the recommendation. Wu et al.~\cite{wu2022selective} define selective fairness task in sequential recommendation where users can flexibly choose which features will be considered sensitive. To satisfy users' diverse fairness demands, they adopt a pre-training and prompt-tuning framework. A traditional recommender system without fair consideration is obtained via pre-training, and diverse fairness needs are satisfied with both task-specific and user-specific prompts using adversarial training. Wang et al.~\cite{wang2022user} present the initial step towards personalized filter bubbles mitigation. While the filter bubble issue~\cite{nguyen2014exploring}  would lead to recommending homogeneous items, it is unreasonable for users to passively accept the recommendation strategy to mitigate such an issue. Their work proposes a new framework called user controllable recommender system which allows users to actively control the mitigation of filter bubbles. Cui et al.~\cite{cui2023controllable} tackle the limitation of existing works that only focus on debiasing pre-defined sensitive features. However, users might be interested in several sensitive groups which are unknown in advance. To solve this, they propose  controllable universal fair representation learning to make the representations fair to all possible sensitive attributes.

\vspace{-1ex}

\subsubsection{Multi-Interest Fairness}

To the best of our knowledge, no studies have yet investigated fairness related to multiple interests. In this section, we review research on multi-interest and suggest potential future directions for fairness in the context of multiple interests. Traditional RS utilize a single embedding to learn user preferences. However, users might have diverse interests that cannot be adequately represented by one embedding. For instance, a user could be interested in sports (e.g., basketball) and art (e.g., painting) simultaneously. One single embedding representing the overall interest could be insufficient to identify these interests and to make corresponding recommendations. Therefore, a single representation would lead to a sub-optimal solution. To mitigate this, researchers use multiple embeddings to represent diverse interests. User $u$ has embedding $\mathbf{Z}_u\in\mathbb{R}^{d \times K^\prime}=\{\mathbf{z}_u^k\}_{k=1,...,K^\prime}$ where $\mathbf{z}_u^k$ denotes user $u$'s $k$-th interest among all $K$ interests. After learning the representations, the relevance score of user $u$ to item $i$ is calculated by $max_{k=1}^{K^\prime}{\mathbf{e}_i}^{T} \mathbf{z}_u^k$ where $\mathbf{e}_i$ is the item representation. The items with the highest scores are recommended to the user.

There have been various ways to learn multi-interest user representations ($\mathbf{Z}_u$). MIND~\cite{li2019multi} represents the initial effort where they extract the interests based on dynamic capsule routing. ComiRec~\cite{cen2020controllable} utilizes self-attention to learn multiple interests based on item interactions. Denote user's interaction as $x_1, ..., x_{nx}$ where $nx$ is the length of interacted items and the representation of $i$'s item in the interaction as $\mathbf{e}_i$. The attention of how much an item is correlated with $k$-th interest is computed with the softmax operations. The user's $k$-th interest is the weighted average of the interacted items based on the calculated attention. These are shown in Eq.~\ref{eq.comirec}.
\begin{equation}
\small 
    a_{k,i}=\frac{\text{exp}(\mathbf{w}^T_k \text{tanh}(\mathbf{W_1}\mathbf{e}_i))}{\sum_j \text{exp}(\mathbf{w}^T_k \text{tanh}(\mathbf{W_1}\mathbf{e}_j))}, \text{ where }
    \mathbf{z_k}=\sum_j a_{k,j}\mathbf{W_2}\mathbf{e}_j.
    \label{eq.comirec}
\end{equation}
Following ComiRec, which considers the item-to-interest relationship, Re4~\cite{zhang2022re4} proposes backward flow to model interest-to-item relationship by adding three regularizations including re-contrast which leverages contrastive learning to learn distinct interest representations, re-attend to ensure that the learned attention correlates to the relevance score for recommendations, and re-construct to highlight that interest representations should reflect the content of representative items. PinnerSage~\cite{pal2020pinnersage} clusters items from the users' interactions with the Ward hierarchical clustering method~\cite{ward1963hierarchical} and uses one representative in each item cluster to represent one of the user's interests. The representative embedding is selected by minimizing the sum of distance with the items in the same cluster. In their work, they find that while the proposed multi-interest strategy improves utility performance, it also increases recommendation diversity. MIP~\cite{shi2022every} also utilizes cluster-based methods to achieve multi-interest. They assign each interest as the latest item representation in each cluster. In addition, rather than treating each interest with uniform importance, they learn the weight to represent the preference over each interest embedding.

There is another line of similar research called disentangled learning. Unlike the multi-interest work that assigns multiple interest embeddings to each user, disentangled learning seperate user embeddings into several sub-embeddings 
that each represent one interest/intention. MacridVAE~\cite{ma2019learning} performs disentanglement at both a macro (i.e., to buy a shirt or a cellphone) and a micro (i.e., the size or the color of the shirt) level based on VAE~\cite{kingma2013auto,rezende2014stochastic}. Macro disentanglement is achieved by learning several prototypes based on users' intentions. Micro disentanglement is realized by magnifying the KL divergence. DisenGCN~\cite{ma2019disentangled} updates the traditional aggregation strategy where nodes gather information from all neighbors uniformly or based on degrees. It aggregates information from related neighbors according to their closeness with the factors obtained from Softmax. For example, sports-related factors will be mainly updated by items like baseball rather than paintings. DGCN~\cite{wang2020disentangled} enhances DisenGCN by applying the distance correlation for factor independence and 
a new aggregation mechanism. It models the distribution over intents for each user-item interaction and iteratively refines the intent-aware interaction graphs and representations.

While multi-interest and disentanglement strategies have demonstrated effectiveness in improving utility performance, most of these studies assume that all users have the same number of interests/intentions. However, users' interests are diverse, and the level of diversity might vary between individuals. Some users might be interested in many aspects, while some users have more specific preferences towards certain categories. Naturally, users with diverse interests should be assigned more interest numbers to capture this diversity. 
Thus, it could be unfair to assign the same number to every user. The impacts of multi-interest or disentanglement for users with varying levels of interest diversity (to the best of our knowledge) have not been investigated, and potential unfairness problems might arise. In addition to the user-side unfairness, investigating item-side fairness is also valuable, as one potential reason for improved performance might be recommending more popular items, thereby increasing popularity bias.

\subsection{Fairness for Item Diversity}

On the one hand, diversifying recommendations helps discover potential user interests to improve user experience, as discussed in Section.~\ref{sec-diversity}. On the other hand, it helps improve the item visibility~\cite{wu2019recent}, especially those unpopular items from small providers that initially have a low opportunity of being recommended. Researchers have highlighted one of the benefits of diversifying is to satisfy the equal market exposure of providers~\cite{wu2022survey}, which naturally connects diversity with a fairness point of view. More specifically, aggregate diversity measures system-level diversity which reflects the systems' ability to recommend less popular or hard-to-find items and thus is relevant to the exposure fairness among providers. Liu et al.~\cite{liu2018personalizing, liu2019personalized} propose two fairness-aware re-ranking extensions called Fairness-Aware Re-ranking (FAR) and Personalized Fairness-Aware Re-ranking (PFAR) based on xQuAD~\cite{santos2010exploiting} which is designed for result diversification. FAR enhances diversity by boosting the scores for items that belong to new providers. Following the general framework in Alg.~\ref{alg.greedy}, they substitute the objective in line 4 into the following:
\begin{equation}
\small 
    v^* = \text{argmax}_{v\in \mathcal{R}_u\setminus \mathcal{S}} P(v|u)+\lambda\sum_{d\in \mathcal{D}}P(d|u)\mathbbm{1}_{v \in d}\prod_{i\in S}\mathbbm{1}_{i\notin d},
    \label{eq.FAR}
\end{equation}
where $\mathbbm{1}$ is the indicator function and $\mathcal{D}$ is the provider set. The first term corresponds to utility performance based on relevance scores, and the second term is to assign an incentive score for the provider that never appears in the existing recommendation list. If none of the recommended items belongs to provider $d$, the second term is effective, Otherwise, the second term equals zero. In this way, it increases the chance of small providers being recommended. PFAR incorporates personalized consideration based on the assumption that users have different tolerance to the level of diversification. They obtain a tolerance score $\tau_u$ based on information entropy and update Eq.~\ref{eq.FAR} by multiplying this weight in the second term with $\lambda$.

Following Liu et al~\cite{liu2018personalizing, liu2019personalized}, Sonboli et al~\cite{sonboli2020opportunistic} also consider fairness-promoting diversity to improve provider fairness. The main idea is to increase the diversity of recommended item list, which benefits the protected class the most. They update the second term in Eq.~\ref{eq.FAR} so that the objective of re-ranking is to find the adjusted recommendation list that maximizes the relevance while maximizing the diversity (i.e., minimizing the similarity). The similarity metric takes into account the item's difference from the existing list and users' tolerance of diversification towards one feature based on information entropy. Specifically, they define a weighted cosine similarity as: 
\begin{equation}
\small 
    \text{wcos}(\mathbf{e}_i, \mathbf{e}_j, \mathbf{t}_u)=\sum_{f}^{|F|}\mathbf{t}_{ut} \mathbf{e}_{if} \times \mathbf{e}_{jf} \frac{1}{\sqrt{\sum_f \mathbf{t}_{uj}\mathbf{e}_{if}^2}\times \sqrt{\sum_j \mathbf{t}_{uj}\mathbf{e}_{jf}^2}},
    \label{eq.wcos}
\end{equation}
where $\mathbf{e}_i$ and $\mathbf{e}_j$ are the representations of two items, $\mathbf{e}_{if}$ is the $f$-th feature in the representation, and $\mathbf{t}_u$ is user's tolerance of diversification. Based on the pair-wise similarity function, the similarity between one item to an existing recommendation list is defined as $\text{sim}(\mathcal{S}, i)=\sum_{i^\prime \in \mathcal{S}}\text{wcos}(\mathbf{e}_i, \mathbf{e}_{i^\prime}, \mathbf{t}_u)$. Compared with FAR and PFAR where the second term loses the boosting effect once the provider has appeared in the recommendation list, Eq.~\ref{eq.wcos} continues work. Additionally, this work provides a more fine-grained personalization considering per feature for users. FA*IR~\cite{zehlike2017fa} enhances diversity to improve group fairness in another way. They create two queues for protected and unprotected items and integrate them to satisfy a probabilistic ranked fairness test. In this way, they ensure that the proportion of protected candidates would remain statistically above a given minimum.

\subsection{Empirical Investigation between Fairness and Diversity}
\label{sec.empirical}

We empirically investigate the relationship between fairness and diversity through experiments on MovieLens dataset\footnote{https://grouplens.org/datasets/movielens/1m/} where the gender attribute and popularity property\footnote{Top $20\%$ items with highest interaction numbers are regarded as popular and the remaining as less-popular.} are regarded as sensitive features in user and item sides, respectively. Due to space limitation, it is infeasible to conduct a comprehensive study on various base models, fairness/diversity measurements and algorithms which requires dedicated efforts in a new survey. We train the base model based on the representative RS matrix factorization (MF)~\cite{rendle2012bpr}, optimize and evaluate according to  specific measurements and methods for improving fairness and diversity. Regularization-based~\cite{yao2017beyond} and reranking-based~\cite{kelly2006enhancing} methods are adopted to enhance fairness and diversity. Fairness is defined as the ratio discrepancy between advantage and disadvantage groups. Aggregate diversity is the total number of different recommended items and we adopt ILD in Eq.~\ref{eq.ild} as individual diversity where the distance function is the cosine distance between pre-trained embeddings. We repeat the experiments three times for each method and report the average evaluation. Note that the learned embeddings in models with regularizations are different from those of vanilla model, this makes individual diversity based on the embedding space incomparable. Therefore, we leave out comparing the individual diversity for fair models. Reranking does not change embeddings and we will include the discussion for the diversity-enhanced methods.
\begin{table}[t]
\centering
\footnotesize
\setlength\tabcolsep{2pt}
\caption{{Empirical relationship between fairness and diversity in RS. All metrics are the higher the better. The best performance of each metric is marked in bold.}}
\label{tab:empirical}
\begin{tabular}{@{}lccccc@{}}
\toprule
\textbf{Model} & \textbf{Utility$\uparrow$} & \textbf{User fairness$\uparrow$} & \textbf{Item fairness$\uparrow$} &  \textbf{Individual diversity$\uparrow$} & \textbf{Aggregate diversity$\uparrow$}\\ \midrule
Vanilla (MF) & \textbf{0.298} & 0.856 & 0.029 & 0.682 & 0.730 \\
+User fairness & 0.255 & \textbf{0.953} & 0.029 & 0.708 & 0.711 \\
+Item fairness & 0.257 & 0.810 & \textbf{0.094} & 0.530 & 0.781 \\
+Individual diversity & 0.272 & 0.857 & 0.058 & \textbf{0.759} & 0.793 \\
+Aggregate diversity & 0.296 & 0.832 & 0.032 & 0.685 & \textbf{0.955} \\ \bottomrule
\end{tabular}
\end{table}
From the results in Table \ref{tab:empirical}, we draw the following observations:
\begin{itemize}[leftmargin=2ex]
    \item User fairness regularization achieves a significant improvement on the user fairness while the other metrics remain at a similar level with vanilla model.
    \item When item fairness improves, the aggregate diversity improves. It indicates that unpopular items are probably recommended due to fairness requirement and thus aggregate diversity is improved.
    \item When individual diversity increases, aggregate diversity also increases since new items are recommended. The item fairness improves since unpopular items have more chance to be recommended based on diversity consideration. Improving individual diversity cannot ensure increasing aggregate diversity or item fairness, but empirically we observe the positive effect.
    \item With reranking to enhance aggregate diversity, the aggregate diversity metric improves significantly while other metrics maintain similarly. There is a slight improvement in individual diversity but it is not as strong as the gain brought by increasing individual diversity to aggregate diversity. Similar case happens for item fairness. The reason for marginal change in other aspects is that few instances need to adjusted for a high aggregate diversity. Therefore, the performance are close to vanilla model. By recommending new items that have never been recommended globally, there is no guarantee on improvement of fairness aspect or individual diversity.

\end{itemize}

\section{Challenges and Opportunities}
\label{sec-challenges}
Researchers have raised the awareness of beyond-utility perspectives to evaluate recommendation including fairness and diversity and have also started the exploration of their intersections. While there exist many challenges, there are also valuable open opportunities for future research directions. 

\begin{itemize}[leftmargin=2ex]
    \item \textbf{Understanding the Relationship between Fairness and Diversity:} 
    This survey delves into the high-level conceptual connections between fairness and diversity and present an initial empirical discussion. However, their practical relationship is intricate and multifaceted. Various metrics exist for evaluating fairness and diversity. It is recommended to conduct extensive empirical studies to determine the feasibility of simultaneously optimizing these metrics or if they inherently conflict. A quantitative analysis of their mathematical relationships is essential.  Additionally, how to incorporate diversity into fairness or vice versa rather than treating them as two goals during the method design is also interesting. Discussions on other types of fairness-diversity intersections beyond user-level and item-level (e.g., single-side vs multi-side) are encouraged.
    \item \textbf{Multi/Many-objective Selection, Optimization and Evaluation:} There are trade-offs between utility and beyond-utility objectives, and also within beyond-utility objectives with diverse metrics. Several research questions need to be addressed \textit{(1) Metric selection: Within various metrics, how to choose the specific ones in practice.} Guidelines for the applications of different fairness and diversity measurements and a thorough theoretical and empirical investigation of the relationship between metrics are encouraged~\cite{smith2023scoping}. \textit{(2) Model optimization: How to balance different objectives, especially when the objectives conflict with each other.} The research field needs to go beyond assuming a single ``best'' model can be obtained due to these inherent trade-offs. Instead, efforts should switch to multi-objective approaches~\cite{wang2023multi,zheng2022survey} 
    that are then evaluated according to their discovered Pareto frontiers. This not only helps better benchmarking and comparison across published works but can provide a suite of non-dominated options/solutions for industry practitioners allowing an increasingly fair decision-making process. 
    \textit{(3) Model evaluations: How to compare and evaluate the model performances when multiple metrics are provided.} It is worth further investigating how to aggregate diverse metrics into one single metric for comparison purposes where the scales and variations of metrics might be different. Rank-based evaluations via the average of the ranks in multiple metrics could avoid the scale issue but cannot be applied in model selection. Addressing these challenges requires dedicated and extensive research efforts.
    \item \textbf{Intersection among Beyond-Utility Perspectives:} While each beyond-utility perspective is often investigated independently, their intersections are of significance in practice. Other intersections beyond fairness-diversity intersection are worth investigating. For example, fairness and explainability are closely connected. Giving the causality for a prediction can expose why certain recommendations, providing insights into the source of bias and potential better ways to mitigate bias~\cite{ge2022explainable, fu2020fairness,medda2023gnnuers}. However, researchers should be cautious when utilizing explanations for fairness where the generated explanations heavily depend on the detailed technique. It is also unclear whether improving fairness related to explainability could ensure fair outcomes.
    \item \textbf{Fairness and Diversity in a Dynamic Setting:} Most works in these two fields focus on static settings where the interactions and user/item profiles remain the same. However, the interactions, user/item profiles are dynamic in practice. It is unclear how the bias and diversity evolve when the distribution of users and items shift. The solutions for static setting can be applied in the latest snapshot which is effective but time-consuming. How to efficiently provide fair or diverse recommendations after the change also needs to be answered. Fair or diverse solutions will make an impact on recommendations and thus influence user's behaviors, what would be the impact in the long run. Furthermore, in these dynamic settings, it would be of interest to investigate how adversarial attacks on RS~\cite{fan2021attacking,deldjoo2021survey} interact with these more responsibly/ethically developed RS.
\end{itemize}
\section{Conclusion}
\label{sec-conclusion}

In this survey, we aim to explore the connections between fairness and diversity in recommender systems. We begin the survey with introductions to preliminaries of recommender systems and relevant concepts on fairness and diversity in recommender systems. After reviewing existing works in fairness and diversity independently, we extend the diversity concept from the item level to include the user level where categorization is provided, including explicit/implicit features, historical preferences (proportionality), fairness needs, and multiple interests (general). With the expanded diversity perspective, we discuss the connections between fairness and diversity from both levels by interpreting fairness works from a diversity point of view. This novel perspective enables a better understanding of existing fairness works and reveals potential future directions. Finally, we discuss the challenges and opportunities with the hope of inspiring future innovations and highlighting the focus on beyond-utility aspects along with their intersections. We hope this survey serves as a valuable resource for future research in recommender systems, particularly in exploring the intersections of fairness and diversity.

\bibliographystyle{ACM-Reference-Format}
\bibliography{references}


\end{document}